\documentclass[fleqn,usenatbib]{mnras}

\usepackage{newtxtext,newtxmath}

\usepackage[T1]{fontenc}
\usepackage{ae,aecompl}

\usepackage{graphicx}	
\usepackage{amsmath}	
\usepackage{amssymb}	

\title[Triple common envelope evolution]{Estimating the outcomes of common envelope evolution in triple stellar systems}

\author[T. A. F. Comerford et al.]{
T. A. F. Comerford$^{1}$\thanks{E-mail: tafc2@cam.ac.uk} and 
R. G. Izzard$^{2}$
\\
$^{1}$Institute of Astronomy, University of Cambridge, Madingley Road, Cambridge CB3 0HA, United Kingdom\\
$^{2}$University of Surrey, Guildford, Surrey GU2 7XH, United Kingdom}

\date{Accepted XXX. Received YYY; in original form ZZZ}

\pubyear{2020}

\begin{document}
\label{firstpage}
\pagerange{\pageref{firstpage}--\pageref{lastpage}}
\maketitle

\begin{abstract}
We present a new model describing the evolution of triple stars which undergo common envelope evolution, using a combination of analytic and numerical techniques. The early stages of evolution are driven by dynamical friction with the envelope, which causes the outer triple orbit to shrink faster than the inner binary. In most cases, this leads to a chaotic dynamical interaction between the three stars, culminating in the ejection of one of the stars from the triple. This ejection and resulting recoil on the remnant binary are sufficient to eject all three stars from the envelope, which expands and dissipates after the stars have escaped.
These results have implications for the properties of post-common envelope triples: they may only exist in cases where the envelope was ejected before the onset of dynamical instability, the likelihood of which depends on the initial binary separation and the envelope structure. In cases where the triple becomes dynamically unstable, the triple does not survive and the envelope dissipates without forming a planetary nebula.
\end{abstract}




\section{Introduction}
Common envelope evolution (CEE) is a phase of multiple stellar evolution during which two or more stars orbit within a shared envelope of gas.
The case of binary common envelope evolution (BCEE) has been an object of study for decades which has produced successful models that match observations of existing binary and single stellar systems.
However, there is no reason to expect that CEE cannot occur in higher-order multiples, particularly those that contain close binaries consisting of compact objects.
While there are several existing prescriptions for determining the outcomes of BCEE, none currently exist for triple common-envelope evolution (TCEE).

BCEE is often the result of unstable Roche lobe overflow (RLOF) in a close binary, during which mass is transferred from one star to its companion faster than the accretor's thermal timescale, meaning it is too fast for the companion to accrete the mass.
The result is the formation of a diffuse gaseous envelope which engulfs both stars.
As the binary continues to evolve inside the envelope, the two cores may merge.
However, if the envelope is sufficiently loosely bound to the cores, it may be ejected \citep{ivanova13}.

Existing prescriptions for BCEE can be tuned by examining the prevalence of stars formed during BCEE mergers, and the separation distribution of post-BCEE binaries.
In this way, it is possible to investigate the underlying physics of CEE and determine how the initial properties of the binary affect its outcomes \citep{demarco17}.

Triple and higher-order stellar systems account for approximately 10 per cent of systems in which the most massive component is a Sun-like star, and the proportion of high-order multiple systems increases for larger stellar masses \citep{moe17}.
Particularly in systems which contain close binaries, is it possible for a triple companion of a binary to overflow its Roche lobe \citep{glanz20}.
If this mass transfer is unstable, it will lead the the formation of a common envelope around the three stars.
A TCE could also arise as a consequence of BCEE in which the extended envelope undergoes unstable RLOF onto the triple companion, although the stability criteria for the triple mean that this can only occur if the envelope substantially expands during BCEE.
In quadruples and higher-order systems, the rate of RLOF may be enhanced due to dynamical interactions which increase the eccentricity of the smaller orbits \citep{hamers19,lidov62,kozai62}.
TCEE has been previously discussed as a potential origin for the system PSR J0337+1715, a triple star consisting of a millisecond pulsar and two white dwarfs \citep{sabach15}.
In their paper, the authors invoke TCEE as a way to remove the envelope of an AGB star, leaving a remnant white dwarf.

Because in this paper we do not distinguish between these two avenues for producing a TCE, we refer to each of the three components of the triple as `stars' even though, in practice, at least one is a stellar core which has decoupled from its envelope.

In this paper, we present our model of TCEE, starting with an overview in \S\ref{sec:overview} before examining the three main stages in detail: inspiral in \S\ref{sec:inspiral}, three-body dynamical interaction in \S\ref{sec:3body}, and ejection in \S\ref{sec:ejeciton}. We discuss these results and provide an algorithm for calculating the outcomes of TCEE in \S\ref{sec:discussion} before concluding with \S\ref{sec:conclusion}.

\section{Qualitative overview}
\label{sec:overview}
In this section, we briefly summarise the stages of standard BCEE and our model for TCEE, before examining each stage in its own section.

BCEE is characterised by three main stages: plunge, inspiral and ejection/merger \citep{ivanova13}.
During the plunge phase, dynamical friction between the stars and the envelope shrinks the binary orbit on a dynamical timescale, and transfers energy and angular momentum into the envelope.
This is a self-limiting process because as the binary separation shrinks, its orbital velocity increases, and the force of dynamical friction weakens.
Eventually, the binary separation stabilises at a much shorter distance than initially (typically $a \sim 10\,R_{\odot}$).
This marks the start of the inspiral phase, in which dynamical friction continues to extract energy and angular momentum from the binary orbit at a lower rate because of the decreased local density and increased envelope rotation velocity.
The timescale is now determined by rates of energy and angular momentum transfer in the envelope, so the binary orbit shrinks on a timescale that is significantly longer than its period.
As the internal energy of the envelope increases, it becomes more tenuous and loosely bound, until it is eventually ejected, leaving a remnant binary and possibly a low-mass disc of bound material.
This sequence can be interrupted at any stage if the binary separation is sufficiently short that one star fills its Roche lobe.
In these cases, tidal forces between the stars are likely to produce a Darwin-like instability, culminating in a merger of the two stars \citep{darwin80}.

The initial phase of TCEE is similar to the plunge phase of BCEE, with dynamical friction acting on both the binary and triple orbits.
The triple orbit always shrinks, but depending on the masses and separations of the stars, the binary orbit may either shrink or grow.
However, because the timescale of binary orbital evolution is always longer than the triple orbit shrinking, it is the change in the triple orbit that drives the system's evolution.
In TCEE, the analogy of a merger between stars is the possibility of the three-body system becoming non-hierarchical, resulting in an unstable dynamical interaction.
While it is possible to produce a binary merger in TCEE, the stricter condition for three-body stability makes unstable encounters more likely.
If the initial triple separation is long enough, and the binary separation both long enough to avoid a binary merger and short enough to avoid a three-body interaction, it is likely that TCEE proceeds similarly to BCEE.
However, the possibility of an unstable three-body encounter permits a new evolutionary channel in which the three stars move on chaotic trajectories in the centre of the envelope.
The general outcome of these encounters is the ejection of one of the stars, with the remaining two forming a binary.
If the ejected star moves sufficiently quickly, it escapes the system entirely and the remaining stars revert to BCEE.
However, if the ejected star cannot escape, either because its velocity is too low or the drag from the envelope too strong, it returns to the centre, prompting another episode of chaotic evolution.
So, assuming that the envelope is eventually ejected, the main unknown is the nature of the central remnant: either a stable triple, or a binary system.

To study this problem we break it into three phases and describe each in a separate section: the plunge and inspiral in \S\,\ref{sec:inspiral}, 3-body interaction in \S\,\ref{sec:3body}, and ejection in \S\,\ref{sec:ejeciton}. 

\section{Inspiral}
\label{sec:inspiral}
During the plunge and inspiral, the evolution of the system is driven by dynamical friction on the stars due to the surrounding envelope.
Typically, the dynamical friction force is of the form,
\begin{equation}
    F = A \left(v^2 + c^2\right)^{-1},
    \label{eq:df-force}
\end{equation}
where $v$ is the relative velocity of the star and surrounding fluid, $c$ is the ambient sound speed, and,
\begin{equation}
    A \approx G^2 M^2 \rho\,,
    \label{eq:df-constant}
\end{equation}
where $G$ is the gravitational constant, $M$ is the mass of the star, and $\rho$ is the ambient envelope density \citep{bondi44, lee11}.
$A$ typically also includes a coefficient of order unity which accounts for geometrical factors.
For stars moving with a constant relative velocity through gas with a uniform density, $F$ always acts in the direction opposite to the velocity, $v$, producing a retarding acceleration on the star.
If the star is accelerating in a direction perpendicular to its motion, the direction of $F$ may acquire a perpendicular component due to the curved wake of the star.
Under the assumption that the length of the Bondi-Hoyle wake is shorter than the relevant orbital separation, we neglect this effect in this paper.

Equation\,\ref{eq:df-force} only holds when the radius of the star is shorter than its Bondi-Hoyle radius,
\begin{equation}
    r_{\textrm{BH}} = \frac{2GM}{v^2 + c^2}\,.
    \label{eq:bh-radius}
\end{equation}
When the radius of the star is longer than $r_\mathrm{BH}$, Eq~\ref{eq:df-force} underestimates the drag force on the star because in this regime the drag is dominated by ram pressure acting on the star's surface.
Enforcing the condition that the stellar radii are shorter than their Bondi-Hoyle radii places constraints on the orbital velocities and stellar radii we consider in this treatment.

\begin{figure}
    \centering
    \includegraphics[width=\columnwidth]{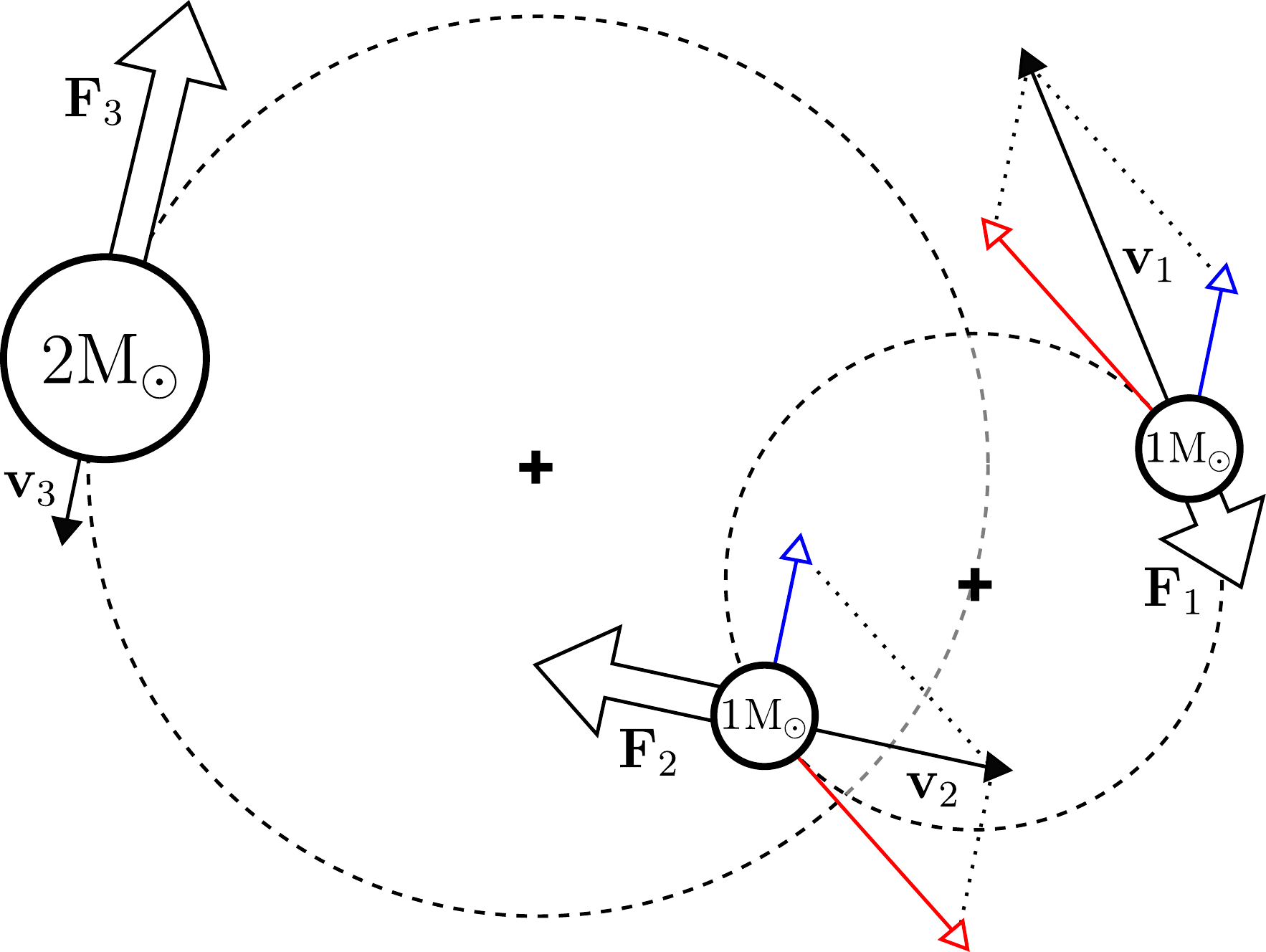}
    \caption{Schematic showing the initial configuration of our triple system. The large and small dashed circles show the triple and binary orbits, respectively, although not to scale. Thin arrows show the stars' velocities in the triple centre-of-mass frame. In the binary, the blue and red arrows with outlined heads show the contributions of the triple and binary orbits to the stars' total velocity, respectively. The thick arrows represent the magnitude and direction of our assumed dynamical friction force on each star.}
    \label{fig:triple-diagram}
\end{figure}

To make the problem more tractable, we make certain assumptions about the triple system. The binary and triple orbits are initially circular and coplanar, and the binary consists of two stars, both of mass $M$, with a triple companion of mass $M_3 = 2M$.
We define the binary and triple semi-major axes as $a_\mathrm{B}$ and $a_\mathrm{T}$, respectively.
Figure~\ref{fig:triple-diagram} shows the physical layout of the system, as well as the stars' velocities in the triple centre-of-mass frame and the dynamical friction force on each of the stars.

\begin{figure}
    \centering
    \includegraphics[width=\columnwidth]{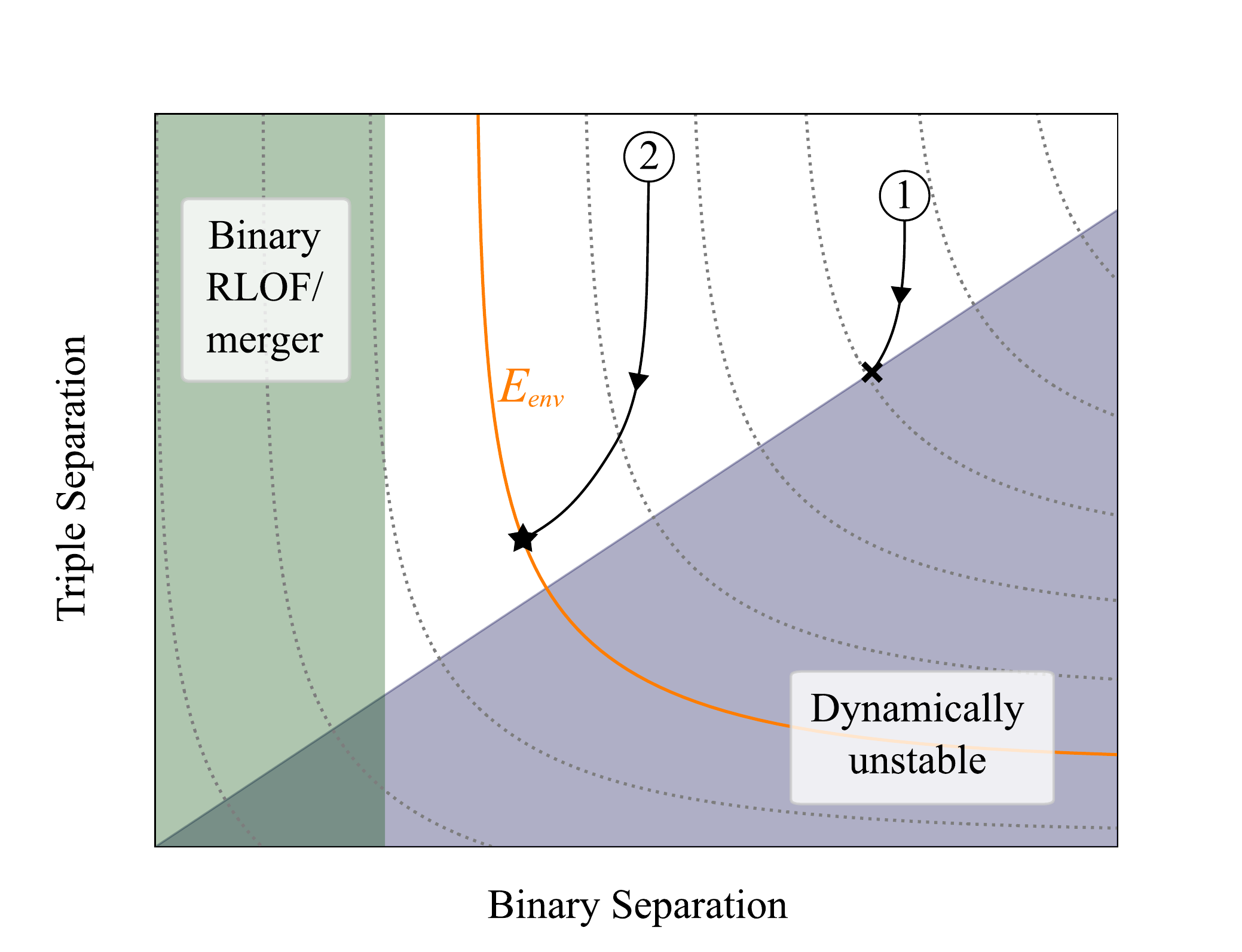}
    \caption{Possible TCEE evolution tracks in the space of binary and triple separation, both plotted on a logarithmic scale.
    The green shaded region shows binary separations at which the binary would merge; the blue shows configurations that are dynamically unstable.
    The dotted lines are contours of equal $E_{\mathrm{env}}$, the energy transferred into the envelope, increasing towards the bottom-left; the orange line highlights one particular value of $E_{\mathrm{env}}$.
    The numbered circles and attached lines show two initial configurations of the orbits and their evolution tracks.
    In case 1, the system starts with relatively small triple separation, and quickly evolves into dynamical instability (indicated with the `$\times$' symbol).
    In case 2, the binary orbit also shrinks, avoiding instability and ejecting the envelope once the energy extracted from the orbits reaches $E_{\mathrm{env}}$ (indicated by the star symbol).
    Note that if the required energy were higher (i.e. a lower contour), the system would approach either dynamical instability or a binary merger.}
    \label{fig:schematic-plot}
\end{figure}

Considering such an `equal-mass' triple introduces symmetries which make determining the evolution more straightforward; namely the binary centre of mass and the triple companion move with equal speeds, through gas with equal density.
As a result, when considering the plunge and inspiral phase, it is not necessary to know the density profile of the envelope to determine the evolution in the $a_\mathrm{B}$-$a_\mathrm{T}$ plane (Fig~\ref{fig:schematic-plot}).

When calculating the effect of dynamical friction on the stars, there is one additional effect to consider: when the separation of the binary is shorter than the Bondi-Hoyle radius of its combined mass, it experiences dynamical friction in a manner similar to a single star of the binary's combined mass \citep{antoni19,comerford19}.
This is important because, owing to the mass scaling in Eq.~\ref{eq:df-constant}, treating the binary as a single star doubles the frictional force it experiences, which increases the magnitude of the torque on the triple orbit.

When calculating the torque on the system due to dynamical friction, it it useful to introduce some new parameters: the orbital velocities of the binary and triple orbits are $v_\mathrm{B}$ and $v_\mathrm{T}$, respectively.
We denote the Mach number of the triple orbit $\mathcal{M} = v_\mathrm{T} / c$, and the ratio of binary and triple velocities $\beta = v_\mathrm{B} / v_\mathrm{T}$.
With our configuration of masses,
\begin{equation}
    v_\mathrm{B} = \sqrt{2GM/a_\mathrm{B}}\,,
\end{equation}
and
\begin{equation}
    v_\mathrm{T} = \sqrt{4GM/a_\mathrm{T}}\,,
\end{equation}
$\beta$ depends only on the ratio of semi-major axes,
\begin{equation}
    \beta = \sqrt{\frac{a_\mathrm{T}}{2a_\mathrm{B}}}\,.
\end{equation}
Finally, $\phi$ is the phase of the binary orbit relative to the triple companion such that when $\phi = 0$, the system is colinear.

Using expressions for the positions and velocities of each of the stars in the binary centre-of-mass and triple centre-of-mass frames, we may use Eq.~\ref{eq:df-force} to determine the resulting torques on each orbit.
A more detailed derivation of the torques appears in Appendix~\ref{app:torques}.
The torque on the binary orbit due to dynamical friction is,
\begin{equation}
    T_\mathrm{B} = \frac{-2Aa_\mathrm{B}}{c^2} \left(
        \frac{\beta \cos (2\phi) + \cos{\phi}}{f_+ \left[ \mathcal{M}^2 f_+^2 + 4 \right]} +
        \frac{\beta \cos (2\phi) - \cos{\phi}}{f_- \left[ \mathcal{M}^2 f_-^2 + 4 \right]}
    \right)
    \label{eq:binary-torque}\,,
\end{equation}
where $f_\pm$ is,
\begin{equation}
    f_\pm = \sqrt{1 + \beta^2 \pm 2\beta \cos \phi}\,.
\end{equation}
Note that this assumes that the binary orbit is smaller than the Bondi-Hoyle radii of each of its stars.
If this is not the case, the true binary torque is smaller than Eq.~\ref{eq:binary-torque}, and the resulting rate of contraction of the binary orbit is overestimated \citep{antoni19, comerford19}.

The triple torque is more complex, because it depends on whether the binary separation is shorter than its own Bondi-Hoyle radius. When $a_\mathrm{B} < r_\mathrm{BH}$, the triple torque is
\begin{equation}
    T_{\mathrm{T}, a_\mathrm{B} < r_\mathrm{BH}} = \frac{ - 16 A  a_\mathrm{T}}{c^2 \left( \mathcal{M}^2 + 4 \right)}
    \label{eq:triple-torque1}\,.
\end{equation}
When the binary separation is larger than $r_\mathrm{BH}$, the triple torque is
\begin{equation}
\begin{split}
    &T_1 = \frac{2Aa_\mathrm{B}}{c^2 f_+ \left( \mathcal{M}^2 f_+^2 + 4 \right)}
    \left( \beta \sin^2 \phi + (2\beta^2 + \cos \phi)(1 + \beta\cos\phi) \right) \\
    &T_2 = \frac{2Aa_\mathrm{B}}{c^2 f_- \left( \mathcal{M}^2 f_-^2 + 4 \right)}
    \left( \beta \sin^2 \phi + (2\beta^2 - \cos \phi)(1 - \beta\cos\phi) \right) \\
    &T_3 = \frac{-16 A \beta^2 a_\mathrm{B}}{c^2 \left( \mathcal{M}^2 + 4 \right)}\\
    &T_{\mathrm{T}, a_\mathrm{B} > r_\mathrm{BH}} = T_1 + T_2 + T_3\end{split}
    \label{eq:triple-torque2}
\end{equation}

Eq~\ref{eq:triple-torque2} suggests that the torque on the triple orbit is never zero.
This is physically unrealistic, as we expect the lost angular momentum to be transferred to the envelope, reducing the velocity discrepancy between the stars and the envelope over time.
If the envelope reaches corotation with the stars, the torque on the triple orbit is zero.
This can be accounted for by subtracting the local envelope rotation velocity from $v_\mathrm{T}$, thus increasing $\beta$.

Calculating the exact value of the envelope's rotational velocity requires knowledge of its density and angular momentum profile.
However, because both components of the triple, i.e. the binary and its companion, are moving through gas with identical density and velocity, we can treat the envelope's rotational velocity at the radius of the stars' orbit as a free parameter, $v_{\mathrm{G}}$.
Realistically, the envelope rotates in the same sense as the triple orbit and is sub-Keplerian, giving the condition $0 \leq v_{\mathrm{G}} \leq v_{\mathrm{T}}$.

\begin{figure*}
    \centering
    \includegraphics[width=\textwidth]{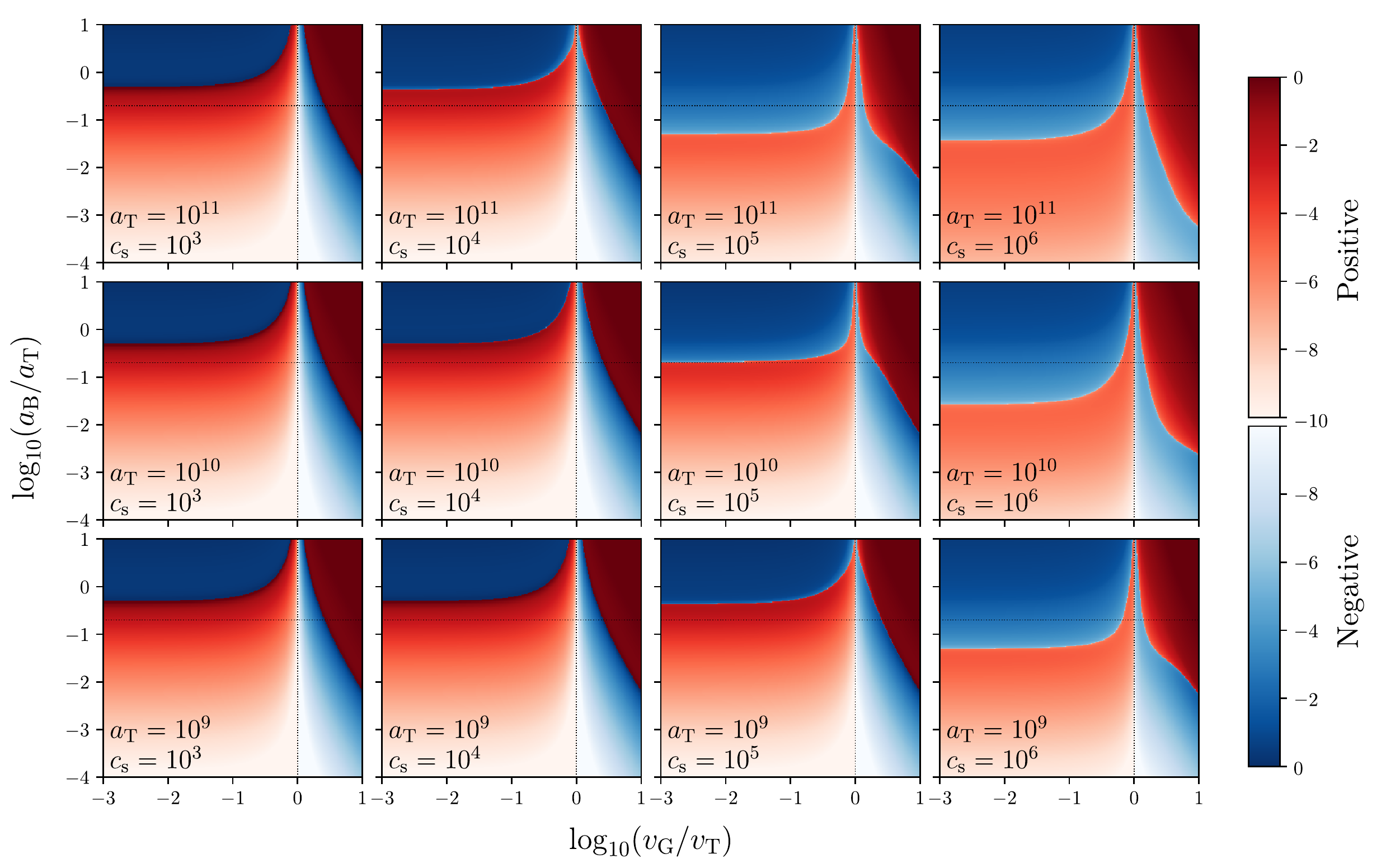}
    \caption{Ratio of torques acting on the binary and triple orbits. Blue (red) regions indicate that the torque on the binary is positive (negative); the torque on the triple orbit is always negative. The horizontal axis is the ratio of the rotation velocity of the envelope, $ v_\mathrm{G}$, to the triple orbital velocity $ v_\mathrm{T}$; the vertical axis is the ratio of binary and triple separations. The dashed black lines delimit the realistic (lower left) region of parameter space: $v_{\mathrm{G}} < v_{\mathrm{T}}$, and $a_{\mathrm{T}} > 5 a_{\mathrm{B}}$. Each column of subplots shows results for a certain value of the sound speed,  $c_{\mathrm{s}}$; each row a different value of $a_{\mathrm{T}}$ (measured in $\mathrm{m~s}^{-1}$ and $\mathrm{m}$, respectively).}
    \label{fig:torque-ratio}
\end{figure*}

Figure~\ref{fig:torque-ratio} shows the ratio of the binary and triple torques over a range of parameters, assuming that the two orbits are prograde and coplanar.
In the physically realistic (lower-left) region of each subplot, the ratio of torques is predominantly positive and always less than unity.
The greatest positive ratio of the two torques in a realistic, stable triple is approximately 0.03.
As a consequence, both orbits shrink, but the triple orbit shrinks faster.
In each subplot, we would expect the system to initially have small $v_{\mathrm{G}}$.
Over time, the ratio of semi-major axes would increase as the triple orbit shrinks, while the ratio of velocities could either increase or decrease as both $v_\mathrm{G}$ and $v_\mathrm{T}$ are increasing.

While rare, there are also some regions of the parameter space in which the orbits experience torques in opposite directions (blue regions in Fig~\ref{fig:torque-ratio}).
In these systems the binary orbit widens while the triple shrinks.
However, even in these cases, the ratio of the magnitudes of the torques is much less than unity, meaning that the shrinking of the triple orbit always dominates the evolution of the system.

In the case of coplanar but retrograde orbits, the size of the torque ratio is the same as prograde orbits, but the sign changes.
The binary orbit in these systems usually widens but, as before, the triple orbit still contracts more quickly.
Eq.~\ref{eq:binary-torque} overestimates the binary torque when the binary separation is shorter than the binary stars' individual Bondi-Hoyle radii. The ratio of triple-to-binary torques exceeds that calculated above, hence the orbital contraction timescales are even more disparate.

If we assume that the ordering of velocity scales is such that $v_\mathrm{B} > v_\mathrm{T} > c$, one can generalise this treatment to other mass ratios, and derive analytic expressions for the binary and triple orbital evolution, allowing for the fact that the two bodies encounter gas with different densities (Appendix~\ref{app:df-mass-ratio}).
The result is a conservative condition for reaching dynamical instability instead of the binary orbit contracting faster than the triple,
which depends on the envelope's density profile. If we assume a radial power-law with an index of -4, the triple should reach dynamical instability regardless of the ratio between the combined binary mass and the mass of the triple companion.

\section{Three-body interaction}
\label{sec:3body}

The results of the previous section show that the likely outcome of the inspiral phase is an unstable three-body interaction.
In this section, we calculate the likely outcomes of these interactions by directly simulating the three-body dynamics using \textsc{rebound} \citep{rein2012}, a library for performing N-body dynamical integration.

Initially, we determine the extent to which the three-body interaction is modified by the continued presence of dynamical friction.
To do this, we calculate the ratio of the triple orbital evolution timescale and the triple orbit period.

Assuming that the stars are sufficiently small that Eq.~\ref{eq:df-force} holds, the timescale for orbital evolution due to dynamical friction is (ignoring constants of order unity),
\begin{equation}
    \tau_{\mathrm{df}} \sim \frac{J_\mathrm{T}}{T_\mathrm{T}} \sim \frac{\frac{GM_\mathrm{T}}{a_{\mathrm{T}}} + c^2}{\rho \sqrt{G^3M_\mathrm{T}a_{\mathrm{T}}}}\,,
\end{equation}
where $J_\mathrm{T}$ is the angular momentum of the triple orbit.
Assuming that $GM_\mathrm{T}/a_{\mathrm{T}} \gg c^2$, i.e. that the triple orbit has a high Mach number\footnote{Note that this assumption overestimates the strength of dynamical friction, meaning that the resulting condition, Eq.~\ref{eq:df-negligible-condition}, is conservative.}, the ratio of this timescale to the triple orbital period is,
\begin{equation}
    \frac{\tau_{\mathrm{df}}}{P_{\mathrm{T}}} \sim \frac{M_\mathrm{T}}{\rho a_{\mathrm{T}}}\,,
\end{equation}
meaning that the condition that three-body encounters are not significantly affected by dynamical friction can be rephrased as the condition that the total mass content within the triple orbit is dominated by the stars, not the gas,
\begin{equation}
    \tau_{\mathrm{df}} \gg P_{\mathrm{T}} \iff M_\mathrm{T} \gg \rho a_{\mathrm{T}}^3\,.
    \label{eq:df-negligible-condition}
\end{equation}
Because, in our model, we assume the total envelope mass is less than that of the stars, the condition Eq.\ref{eq:df-negligible-condition} is automatically satisfied, and the effect of dynamical friction during the unstable three-body dynamics is negligible.

To determine the outcome of the three-body interactions, we simulate systems with an initial ratio of separations $a_\mathrm{T} / a_\mathrm{B} = 2.5$. This choice of initial separations means that the systems start reasonably hierarchically, but chaotic evolution begins reasonably promptly, lowering the computational cost.

We include the effect of dynamical friction on the stars' trajectories using Eq.~\ref{eq:df-force}, which is multiplied by a parameter that   represents the local gas density.
Because we do not know the density, we treat this as a free parameter, and vary it between $10^{-4}$ and $10^{-8}$.
Since the strength of dynamical friction is proportional to the gas density, this parameter is equal to the density when expressed in units of $M a_\mathrm{B}^{-3}$.

If the gas density is higher than this, there are cases where it is sufficient to completely halt the triple orbit, placing stars on direct collision trajectories.
For coplanar orbits, the only one additional parameter needed to describe the system's initial configurations is the difference in phase between the two orbits.
Simulating systems with a uniform distribution of orbital phase separation yields the probability that any particular star is ejected, along with the distribution of ejection velocities and separation of the remaining binary.

Statistical studies of the unstable three-body problem show that the outcomes depend on the initial energy and angular momentum of the system \citep{stone2019}.
The energy of the system is fully determined by the initial binary and triple separations, while the total angular momentum also depends on the orientation of the two orbits (i.e. their mutual inclination).
Angular momentum is maximised when the orbits are aligned in the same sense (prograde), and minimised when they are in the opposite sense (retrograde).
However, because the orbital angular momentum increases with mass and separation, the triple orbit is the dominant contribution to the system's total angular momentum, and the effect of varying the binary inclination is small.

Under the assumption that the extremal angular momenta bound the distributions of outcomes, we simulate both prograde and retrograde systems. The output parameters of interest are which of the the three stars is ejected, its velocity at infinity, and the semimajor axis and eccentricity of the remaining binary.

We label a triple as unbound when there exists a combination of stars where two are bound into a binary, and the third is unbound, at a distance of at least five times the binary semi-major axis.
As soon as this condition is reached, the simulation is ended, and the ejection velocity at infinity is calculated from the total energy of the (unbound) triple orbit.

\begin{figure*}
    \centering
    \includegraphics[width=\textwidth]{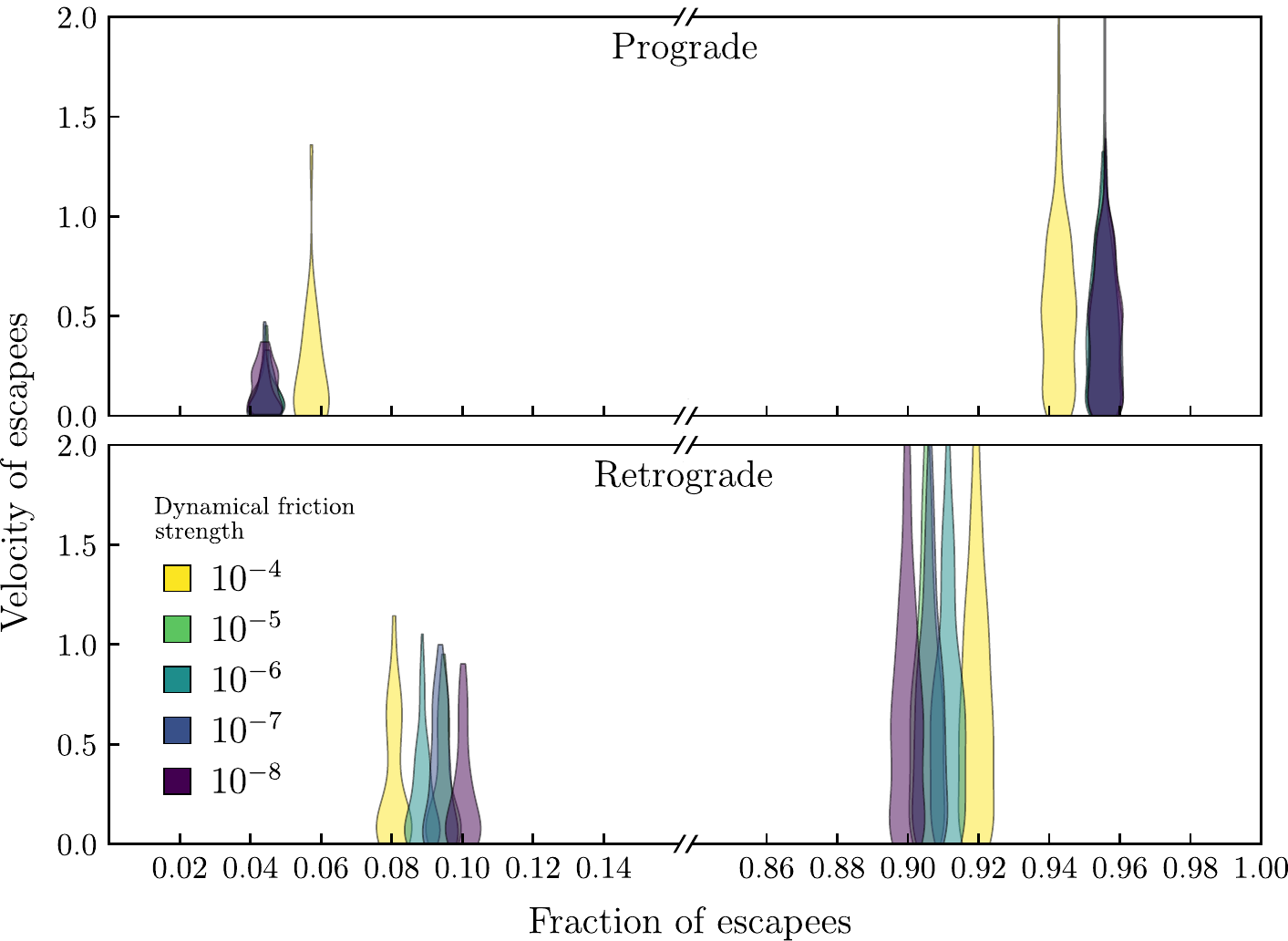}
    \caption{Violin plots showing distributions of outcomes of 3-body interactions for initially prograde and retrograde triples, with velocities at infinite distance expressed in units where the initial binary orbital velocity is 1. The width of each trace is proportional to the number of escapees at the corresponding velocity. Uncertainties in the horizontal direction are roughly 0.01. Points on the left show cases where the more massive star was ejected; points on the right represent cases where an unequal-mass binary was formed, ejecting one of the less massive stars. The colour denotes the strength of the dynamical friction term; note that for prograde triples, all simulations lie at the same escape fraction, apart from those with the strongest dynamical friction (in yellow). Retrograde triples were able to eject a small number of stars at up to 3 times the binary orbital velocity; we have truncated this figure at $v=2$ for clarity. The distributions are naturally truncated at $v=0$ as the `escapees' would otherwise remain bound to the binary.}
    \label{fig:3-body-esc}
\end{figure*}

The resulting velocities and probabilities are shown in Figure~\ref{fig:3-body-esc}.
It is apparent that in all cases, ejection of the most massive star is much less likely, averaging between 4 and 10 per cent; this is an example of mass segregation, in which the most massive components of a gravitationally bound system migrate towards its centre.
We find that when compared to prograde triples, the ejection of the most massive star is roughly twice as likely in retrograde triples, although the ejection probability has a much stronger dependence on the choice of dynamical friction parameter than in prograde triples.
This is probably because in retrograde triples, the components of the binary frequently move very slowly relative to the centre of mass, resulting in comparatively large dynamical friction forces.
Additionally, in absence of dynamical friction, a retrograde triple is more dynamically stable than a prograde triple with otherwise identical parameters.
This allows dynamical friction a longer time to act on the stars, increasing the deviation from pure three-body gravity.
On the other hand, when the relative strength of dynamical friction is less than $10^{-4}$, the prograde systems all have almost identical ejection probabilities and velocity distributions.
When compared to the prograde systems, there is a tail of high-velocity ejections observed in the initially retrograde simulations.
This can be explained by comparing the total angular momentum of the two scenarios: when the triple is initially retrograde, the system has less angular momentum, allowing for high-eccentricity close encounters between stars, during which, interactions with the third star can produce particularly fast ejections.

When the most massive component is ejected, it is generally at a lower velocity, even lower than would be expected if ejections occurred with some characteristic impulse (which would produce velocities inversely proportional to mass).
A substantial fraction of escapees have small excess velocities, so are only marginally unbound.
This is to be expected if the three-body encounter is treated as a series of scrambles\footnote{This is the terminology in \cite{stone2019}, in which it refers to a temporary disordered configuration with no hierarchy.}, which eject one component according to a distribution of velocities.
If that distribution peaks below the local escape velocity, the majority of ejections will be temporary and will eventually result in another scramble.
Thus, the stars which manage to escape are ejected from scrambles with velocities at the high end of the distribution, truncated at the escape velocity which results in the star's velocity approaching zero at infinity.

\section{Ejection}
Provided no mergers or collisions occur during the three-body interaction, we are left with an escaping single star and a recoiling binary in the centre of the envelope.
The evolution of the system from this point is highly uncertain, mainly due to the role of gas pressure in the central regions of the envelope.
Given the complexity of the problem, we do not perform a full treatment of the evolution of the system in this section, rather we estimate what the likely outcomes are.

\label{sec:ejeciton}
\subsection{Predictions using a power-law envelope density profile}
Assuming that the pressure in the envelope approximates hydrostatic equilibrium at the moment that the three-body interaction resolves, the temperature and density profiles in the envelope are such that the force of gravity (almost entirely due to the stars) is balanced by the gas pressure.
As the stars recoil from the centre, they leave behind a region containing gas at high pressure, but lacking the weight needed to contain it.
The consequence of this is an expanding bubble of hot material which does work to raise the overlying envelope and propagates outwards at the local sound speed.

The effect on the stars now depends on their Mach number. If the stars are supersonic, they always lie ahead of the shock, and the envelope material they encounter is not drastically different from the profile during the inspiral and 3-body interaction.
In this case, we may calculate the trajectories of the stars by numerically integrating their motion, including dynamical friction from the as-yet undisturbed envelope.
If the stars are subsonic, however, it is possible that the expanding bubble overtakes them.
It is unlikely that the shock front could propagate too far ahead of the stars, as their gravity aids the remaining envelope to confine the hot bubble.
In this case, it is probably not accurate to describe this stage of TCEE as `ejection', because the escaping stars may retain some of the envelope.

Hydrodynamical simulations of common envelope evolution show that after the inspiral phase, the density profile in the envelope scales as roughly $\rho \propto r^{-4}$ \citep{reichardt19}. Assuming hydrostatic equilibrium in which the the envelope's contribution to gravitational potential is negligible, the corresponding pressure profile is, 
\begin{equation}
    p(r) = \frac{G M_\mathrm{T} \rho_0 r_0^4}{5 r^5},
    \label{eq:pressure-profile}
\end{equation}
where $\rho_0$ is the density at radius $r_0$ and $M_\mathrm{T}$ is the combined mass of all three stars.
The sound speed can then be estimated as a function of radius,
\begin{equation}
    c(r) = \sqrt{\frac{p(r)}{\rho(r)}} = \sqrt{\frac{G M_\mathrm{T} r_0^8}{5 r^9}} \propto r^{-9/2}\,.
    \label{eq:sound-speed}
\end{equation}
Note that since, especially in the inner regions, the envelope is also rotationally supported, Eq.~\ref{eq:pressure-profile} overestimates the gas pressure, and Eq.~\ref{eq:sound-speed} overestimates the local sound speed.

This density profile also allows us to determine the total envelope mass, assuming it is truncated at some small radius to avoid a singularity. Without loss of generality, we can set the inner truncation radius to $r_0$ to find,
\begin{equation}
    M_\mathrm{env} \approx 4 \pi \int_{r_0}^{\infty} \rho(r) r^2 dr = 4 \pi \rho_0 r_0^3\,.
\end{equation}
Now, for a given envelope mass and truncation radius, we can directly determine the density as a function of radius, and integrate stellar trajectories that include the effect of dynamical friction.
It is reasonable to set the truncation radius to the size of the triple orbit as the gas density would be expected to peak in the vicinity of the stars, and any gas inside the triple orbit has no effect on the system's future evolution because its mass is negligible when compared to the stars' masses.

In the event that an escaping star of mass $M_\mathrm{e}$ is ejected at exactly the local escape velocity and does not experience dynamical friction, its velocity relative to the centre of mass is,
\begin{equation}
    v_\mathrm{esc}(r) = \frac{M_\mathrm{T}}{M_\mathrm{e}}\sqrt{\frac{2GM_\mathrm{T}}{r}} \propto r^{-1/2}\,.
\end{equation}
Note that this is an underestimate of the star's velocity. If it escapes its velocity is higher by definition, and the results of \S\ref{sec:3body} show that the velocity is usually substantially higher. In these cases, $v(r)$ is always greater than predicted by the power law $v \propto r^{-1/2}$, meaning that the strength of dynamical friction is overestimated.
Therefore, the Mach number of this marginally-unbound star is,
\begin{equation}
    \mathcal{M}_\mathrm{e}(r) = \frac{v_\mathrm{esc}(r)}{c(r)} \propto r^{4}\,,
\end{equation}
which is to say that if the star escaping is initially supersonic, it remains so unless dynamical friction is sufficient to recapture it.

\subsection{Results using a calculated polytropic envelope}
A more physically-motivated model of the common envelope pressure and density profiles is a polytrope surrounding a point mass.
In the envelope, pressure and density are related by,
\begin{equation}
    P = K \rho^{1 + 1/n},
\end{equation}
where $K$ is a constant which can be expressed in terms of the central density and pressure of the envelope, and $n$ is the polytropic index - here taken to be 3/2 as in a convective envelope.
We may then integrate the equation for hydrostatic equilibrium to determine the envelope's structure for a given central pressure, density and core mass.
Here, `core' refers to the region containing all three stars, which is assumed to be sufficiently small that the envelope remains spherically symmetric.
The core mass is then just the sum of the three stars' masses - here, $4\mathrm{M}_\odot$.
Using this method, we create a set of envelope models parameterised by the envelope radius (in the range $10^{10} \text{ to } 10^{13} \mathrm{~m}$) and mass relative to the cores (in the range $10^{-4} \text{ to } 10^{1}$).

Given an initial relative separation and velocity, the trajectories of the disrupted triple can then be determined, using the density and sound speed profiles of the envelope.
We specify the separation of the remaining binary to be $1/5$ of initial distance between it and the single escapee, matching the conditions at the end of \S\ref{sec:3body}, to account for the reduced dynamical friction when the binary separation is larger than its Bondi-Hoyle radius.

At any given position within the envelope, we define the modified escape velocity, $v_\mathrm{esc,mod}$, as the minimum velocity required for an escaping star to become unbound from the envelope after accounting for dynamical friction.
The strength of the friction is then measured by the quantity,
\begin{equation}
    \Delta(r) = \frac{v_\mathrm{esc, mod}(r) - v_\mathrm{esc}(r)}{v_\mathrm{esc}(r)},
\end{equation}
where $v_\mathrm{esc, mod}(r)$ and $v_\mathrm{esc}(r)$ are the relevant escape velocities at a radius $r$ from the centre of mass.
$\Delta$ is zero in absence of dynamical friction, and $\gtrsim 1$ in cases where friction significantly impedes the stars' escape.

Setting the envelope outer radius to $10^{12} \mathrm{m}$, we perform a series of simulations with varying envelope mass and initial escapee-binary separation.
The envelope mass is expressed as a fraction of the mass of the cores, between $10^{-4}$ and $1$ and the range of initial separations spans $10^{9}$ to $10^{13} \mathrm{m}$.

\begin{figure*}
    \centering
    \includegraphics[width=\textwidth]{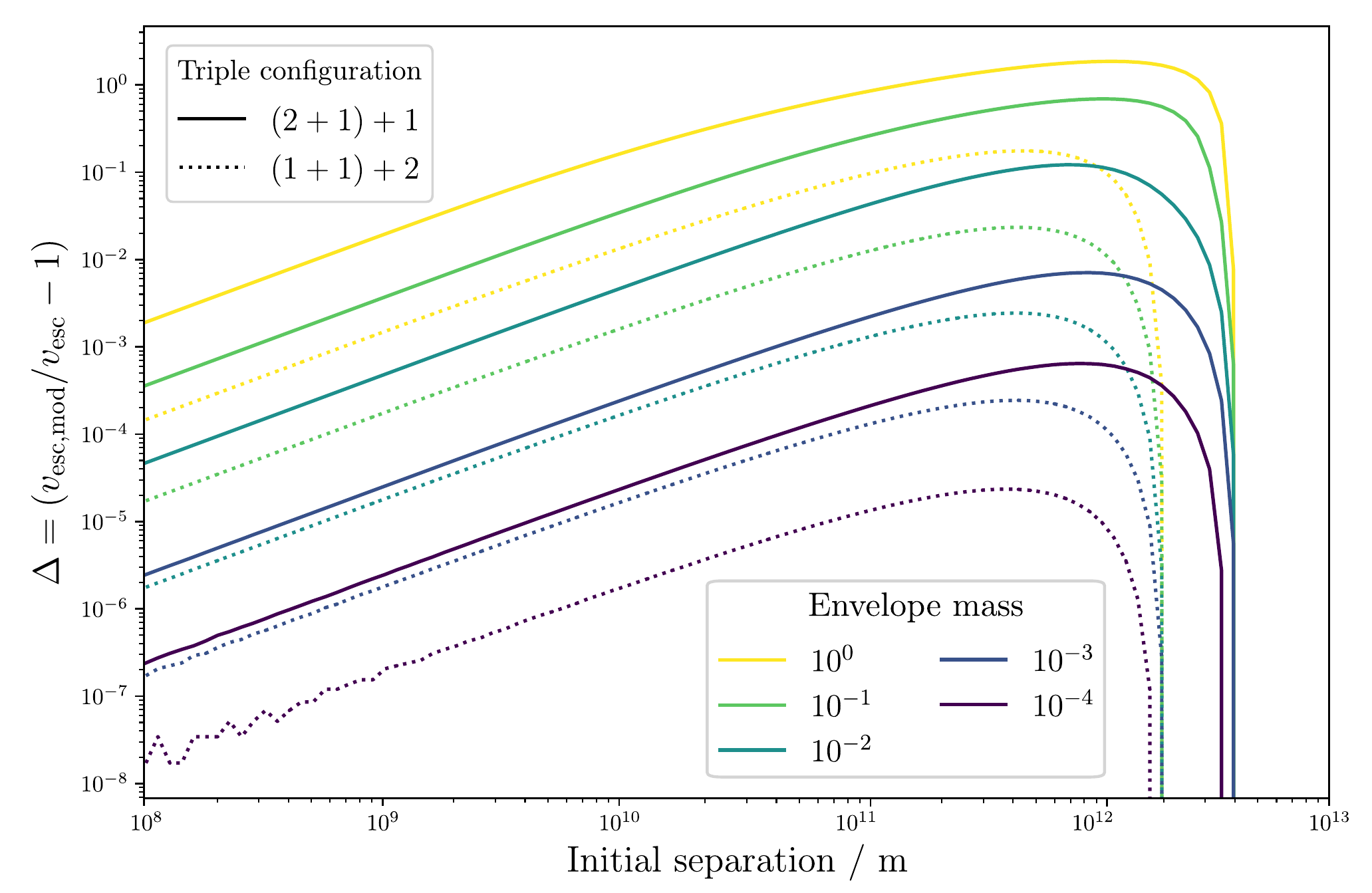}
    \caption{Plot showing the dependence of $v_\mathrm{esc,mod}$ on the initial separation, envelope mass, and triple configuration. Solid lines show disrupted triples where the binary contains a $2 M_\odot$ and a $1 M_\odot$ star, with a $1 M_\odot$ escapee; dotted lines represent configurations with an equal-mass $1 M_\odot - 1 M_\odot$ binary, and a $2 M_\odot$ escapee, which is much rarer, as shown in Section \ref{sec:3body}. While the envelope radius is $10^{12} \mathrm{m}$, the initial separation may be up to twice this length, although the effect of friction is negligible here. Departures from smooth curves at low $\Delta$ are numerical in origin.}
    \label{fig:modified-vesc}
\end{figure*}

The results of our simulations are shown in Figure~\ref{fig:modified-vesc}, demonstrating trends in $\Delta$ that depend on the initial separation, $r_0$, the envelope mass, and the triple configuration.
The relative strength of dynamical friction is greatest in ejections that occur with very long distances between the binary and escapee because $v_\mathrm{esc}$ is much lower in these systems.
When $r_0$ is roughly the radius of the envelope, escaping stars only experience dynamical friction for a short time, meaning $\Delta$ begins to decrease toward zero here.
The differing values of $r_0$ at which $\Delta = 0$ are a consequence of the triple configuration. If the escapee is less massive than the binary, $r_0$ must be longer to ensure that all components are outside the envelope.
When the initial separation is low, $\Delta \propto r_0$, although we have not been able to physically motivate this result.

Unsurprisingly, more massive envelopes present a more effective obstacle to stellar escape, and $\Delta$ is approximately proportional to $M_\mathrm{env}$ in the mass range of our simulations\footnote{Note that the proportionality is not exact, as shown by the variable vertical spacing of the lines.}.

Finally, $\Delta$ depends fairly strongly on which star escapes the triple in this unequal-mass scenario.
Our initial $2 + 1 + 1 M_\odot$ triple may fragment in two different ways, when either a $2 M_\odot$ star escapes leaving an equal-mass binary, or a $1 M_\odot$ star escapes, leaving the unequal-mass $2 + 1 M_\odot$ binary. Section \ref{sec:3body} shows that the latter is more common by around an order of magnitude.
In the unequal-mass cases, the more massive component has a lower initial velocity and distance from the centre of mass.
These two effects combine to strengthen the effect of dynamical friction on that component, which provides the dominant contribution to $\Delta$.
Hence the value of $\Delta$ in unequal-mass triples is higher by roughly an order of magnitude, when compared to the equal-mass `(1+1)+2' triples.

With these results, we may now estimate the magnitude of the impact of dynamical friction on triples which are dissociated in a 3-body encounter.
By writing $v_\mathrm{esc}$ in terms of $v_\mathrm{B}$ using $r_0 = 5 a_\mathrm{B}$, we can derive a relation between $\Delta$ and the excess velocities computed in Section \ref{sec:3body},
\begin{equation}
    \Delta = \sqrt{1 + \frac{5}{4}\left(\frac{v_{\infty}}{v_\mathrm{B}}\right)^2} - 1,
\end{equation}
where the fraction $v_\infty/v_\mathrm{B}$ corresponds to the vertical scale in Figure~\ref{fig:3-body-esc}.

\begin{figure}
    \centering
    \includegraphics[width=\columnwidth]{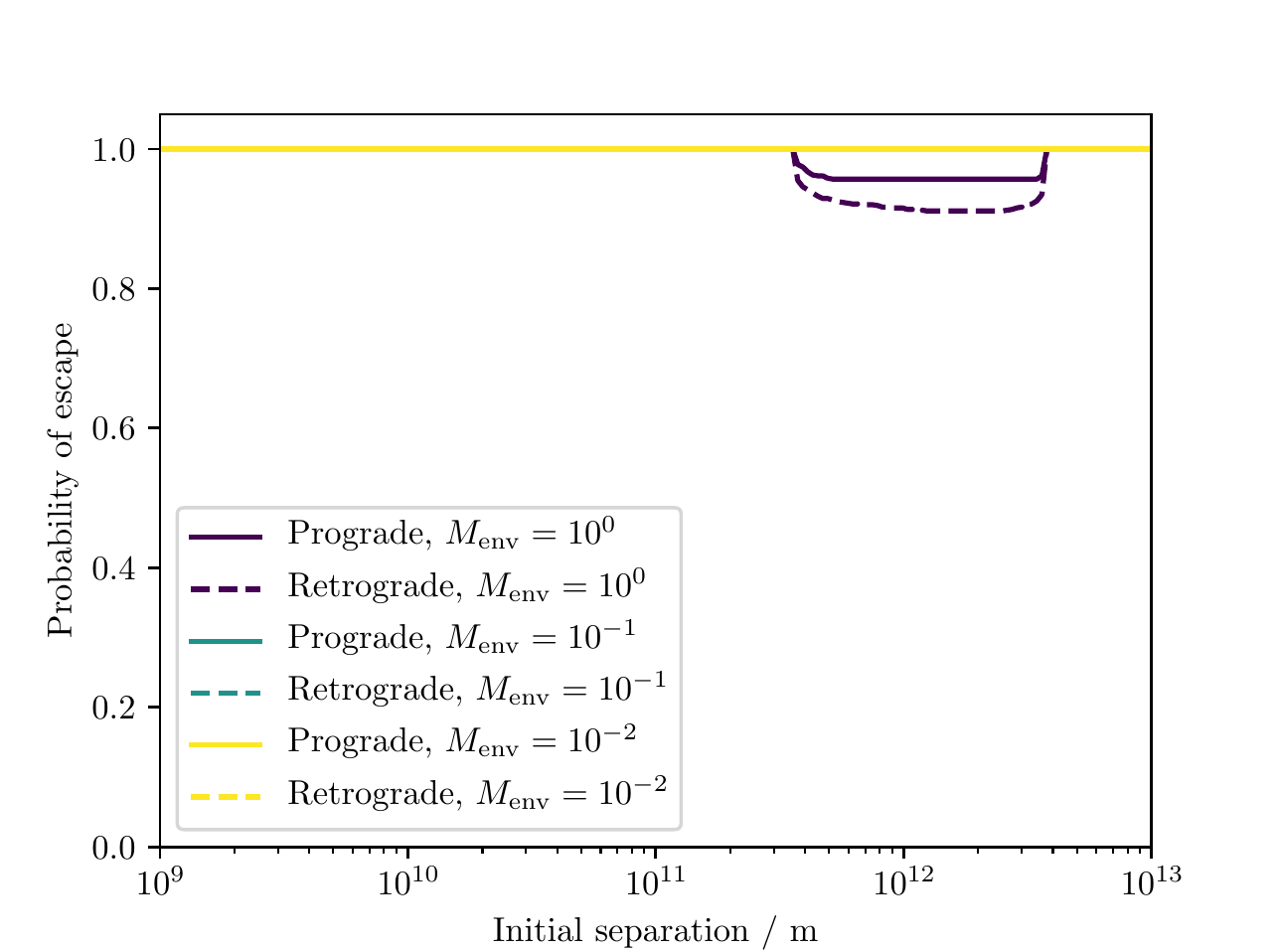}
    \caption{Probability of escape after a three-body encounter and dynamical friction, shown as a function of the initial separation, envelope mass, and triple orientation. Except for initial separations $\approx 10^{12} \mathrm{m}$, all of the lines are overlaid at a probability of 1. Escape is certain, except in cases with very massive envelopes and large initial separations.}
    \label{fig:p-escape}
\end{figure}

For a given envelope mass and triple initial orientation (prograde or retrograde), we may now calculate the probability of escape as a function of separation immediately after the three-body encounter, using the distributions of excess velocities from \S\ref{sec:3body}.
We convert the distributions of $v$ into a distribution of $\Delta$, and then count the number of simulations with $\Delta$ greater than the derived value for the specified parameters.
These distributions (see Figure~\ref{fig:p-escape}) show that in almost all cases, the ejection velocity after the three-body encounter is sufficient to overcome dynamical friction, and allow the stars to escape the envelope.
We find that the envelope is rarely able to recapture stars, and then only when it is particularly massive and the initial separation is similar to the size of the envelope. 

\subsection{Fate of the envelope}
Using the envelope profiles from the previous section, we can determine whether the envelope remains bound after the stars have escaped.
We do so under the assumption that the stars' escape is much faster than the dynamical time of just the envelope, and neglect energy injected into the envelope by the escaping stars.
Now, using the density and pressure profiles, we explicitly calculate the potential and internal energy throughout the envelope.
We say that a layer is unbound if its internal energy is greater in magnitude than its gravitational potential energy, and all layers above it are also unbound (whether the outermost layer is unbound depends on its energy only), i.e. that the envelope is stable against Jeans collapse.

We find that all our polytropic-envelope models are unbound according to this condition.
As a result, the final state of triple common-envelope evolution consists of a binary and a single star, moving away in opposite directions with velocities in the range of 10s to $100\text{s km s}^{-1}$, fast enough that the triple is unbound. The common envelope then dissipates.
The escaping stars may be detectable in large surveys such as Gaia, appearing as a single star and a close binary with an apparent common origin.
Since the the envelope's only source of energy is its own internal energy, it will not form a planetary nebula, as these depend on a central hot source to ionise the gas.

\section{Discussion}
\label{sec:discussion}

\subsection{Approximating outcomes of TCEE}
Using the results of the previous sections, we construct a method to quickly approximate the outcome of triple common envelope evolution.
During the initial phase, the triple orbit shrinks, while the binary remains mostly unaffected as it evolves on much longer timescales.
By analogy with BCEE, the evolution of the envelope may be estimated using one of the several prescriptions, say the $\alpha$ (energy) or $\lambda$ (angular momentum) prescription \citep{webbink84, nelemans05}.
If the envelope is ejected, the triple continues to evolve with its new triple separation, and TCEE ends.
If the envelope is not ejected before the triple becomes unstable, we move on to the next phase of TCEE, the 3-body interaction.

In this phase of TCEE, the triple undergoes chaotic evolution, culminating in the ejection of one of its components.
The probability of ejection for each star depends on the masses and initial configuration of the triple, from which less massive components are more likely to be ejected.
The ejection velocity is likely to substantially exceed the triple's escape velocity, by an amount of similar order to the initial binary orbital velocity.

After one of the components is ejected, both it and the remnant binary escape the surrounding envelope.
In cases where the envelope is less massive than the stars, escape is almost certain.

These predictions have significant consequences for the prevalence of close post-common envelope triples.
While our model does not preclude their existence, it implies that their envelope must have been ejected before the system reached dynamical instability.

\subsection{Limitations and future work}

The model presented in this paper is an overview of TCEE where we have relied on various simplifying assumptions involving the masses of the stars and neglected certain forces and effects.
Where possible, these assumptions have been justified, but there is still scope for more detailed investigations in each of the three phases that we have described.
Of particular interest would be hydrodynamical simulations, which would be able to study the detailed effects of dynamical friction, the envelope response, and departures from spherical symmetry that we have neglected here.
These would also test our assertion that after the three-body ejection, the stars move sufficiently quickly that the envelope structure is not significantly modified during their escape.

Another important refinement would be a more complete investigation of the outcomes of the unstable three-body interaction.
While an implementation of the algorithm above could simply include simulating the three-body system, it would be beneficial to have access to a wider range of mass ratios and configurations.
Alternatively, methods exist to explicitly calculate these probabilities \citep{stone2019}, in addition to novel techniques using machine learning \citep{breen19}.

One particularly important effect that we have neglected is the possibility of interactions between the stars, especially within the binary.
In BCEE, one possible consequence is RLOF on one of the central stars onto the other, often culminating in a merger of the two cores.
While RLOF is less likely in TCEE, there is still the possibility of collisions between the stars during the unstable three-body stage, especially if any of the stars are still on the main sequence.
By analogy with BCEE, such a collision would be likely to release large amounts of energy, potentially unbinding the envelope.
Regardless of the interaction with the envelope, the merging of two stars precludes the possibility of a future chaotic interaction, as the dynamics revert to the stable two-body problem.

Another potentially significant source of energy is the gravitational potential energy released during accretion of the envelope onto the stars.
This could be particularly important if the stars are compact objects, i.e. white dwarfs, neutron stars, or black holes.
In cases with significant accretion, circumstellar accretion disks may also produce jets, which interact with the envelope \citep{schreier19}.

While we have restricted our study to bound triples, our results are similar to those in \cite{davies98}, in which the authors consider the outcomes of collisions between close binary stars and red giants.
This scenario produces two different evolutionary pathways: in the first, the binary passes straight through the envelope, ejecting the giant's core in the process.
In the second pathway, the binary is captured by the envelope, leading to TCEE.
Their description of TCEE generally agrees with our results; the triple orbit shrinks due to friction with the envelope, leading to an unstable three-body interaction in which one component is ejected from the system.
Our results differ because they found no cases in which the surviving binary was also ejected from the envelope; the TCE always became a BCE after one star was ejected.

There are two differences between our work and theirs which may explain this discrepancy: methodology and initial configuration.
Our methods have involved relatively straightforward integration of the stars' trajectories, using simple assumptions about the envelope structure and orbital evolution.
In \cite{davies98}, the authors also use direct integration, but then compare their results with smoothed particle hydrodynamics (SPH) simulations.
Using SPH means that their simulations better capture the response of the envelope, which may produce sufficient extra friction on the surviving binary that it is retained.
The other substantial difference is the context of the study.
We have focussed on TCE as a stage of triple stellar evolution, which has two important consequences for our initial configuration: all three stars are initially bound on circular orbits, and that the envelope is centred on the triple's centre of mass.
In contrast, \cite{davies98} sought to describe the outcomes of collisions between binaries and single giants, meaning that their systems are initially unbound, and that the envelope is centred on one of the stars.
While all three stages of TCEE depend on the initial parameters, the outcomes of the three-body interaction are likely to determine the overall evolution of the system.

We do not believe our assumptions about the initial state of the system, namely having circular orbits with the envelope's centre coincident with the system's centre of mass, significantly affect our conclusions in most cases.
When the envelope is off-centre, as is expected at the onset of CEE, the torque on the triple orbit differs from that calculated in \S\ref{sec:inspiral} because one component of the triple orbit encounters higher-density gas at a lower relative velocity than we assume, while the other encounters less-dense gas at a higher relative velocity.
As a result, the former component experiences stronger dynamical friction than the latter.
As the triple orbit contracts, the distance between the triple and the envelope centres of mass decreases, reducing the disparity in gas density and velocity encountered by the stars.
Eventually, the envelope's centre of mass aligns with the triple centre of mass, as in our original treatment above.

\section{Conclusions}
\label{sec:conclusion}
In this paper, we have presented our model of triple common-envelope evolution, during which a triple-star system progresses through three main stages.
In the first stage, the orbits shrink because of dynamical friction between the stars and the common envelope.
In almost all cases, the triple orbit contracts more quickly than the binary orbit, meaning that the triple eventually becomes dynamically unstable.
If the triple reaches instability, it goes through phase of chaotic three-body dynamical evolution during which the effect of friction with the envelope is negligible.
The outcome of the three-body encounter is the ejection of one of the stars while the remaining two form a binary.
If the stars have unequal masses, the less massive stars are more likely to be ejected singly.

Because the stellar velocities in the three-body encounter are comparatively high, we find that dynamical friction from the envelope is very unlikely to impede the escape of the stars.
Consequently, the likely final result of triple common-envelope evolution is the unbinding of the triple, leaving an escaping binary and single star, while the remaining envelope expands and dissipates.
Using existing common-envelope prescriptions, e.g.~the $\alpha$-prescription for energy or the $\gamma$-prescription for angular momentum, one may determine whether the envelope is ejected before the onset of dynamical instability.
In these cases, the stable triple remains while the envelope dissipates.
This method may be used to approximate the outcomes of common-envelope evolution in triple, and higher-order, stellar-evolution codes.

\section*{Acknowledgements}
We thank the referees for their comments, which helped us to improve this paper.
TAFC thanks the UK Science and Technology Facilities Council for his studentship. RGI thanks the STFC for funding his Rutherford fellowship under grant ST/L003910/1.
Simulations in this paper made use of the REBOUND code which is freely available at http://github.com/hannorein/rebound. We thank Melvyn Davies, Nathan Leigh, Elliot Lynch,   Zephyr Penoyre, Andrew Sellek and Christopher Tout for useful discussions.

\section*{Data availability}
Data available on request.




\bibliographystyle{mnras}
\bibliography{bibliography} 



\appendix

\section{Deriving torques on the binary and triple orbits}
\label{app:torques}

With the above definitions of orbital parameters, we may derive expressions for the orbital velocity of the two components of the binary, in the non-rotating triple centre-of-mass frame,

\begin{equation}
    \mathbf{v}_1 = \frac{v_\mathrm{T}}{2} \begin{pmatrix} \beta \sin \phi \\ 1 + \beta \cos \phi \end{pmatrix},
    \label{eq:v1}
\end{equation}
and
\begin{equation}
    \mathbf{v}_2 = \frac{v_\mathrm{T}}{2} \begin{pmatrix} - \beta \sin \phi \\ 1 - \beta \cos \phi \end{pmatrix}\,.
    \label{eq:v2}
\end{equation}
In the binary centre-of-mass reference frame, the positions of the stars are
\begin{equation}
    \mathbf{r}_1 = - \mathbf{r}_2 = \frac{a_\mathrm{B}}{2} \begin{pmatrix} \cos \phi \\ \sin \phi \end{pmatrix}\,,
    \label{eq:r1}
\end{equation}
respectively.
Note that because they are measured in different frames of reference, $\mathbf{v}_1$ and $\mathbf{v}_2$ are not the time derivatives of $\mathbf{r}_1$ and $\mathbf{r}_2$.
This has no effect on the following calculation of the torque, since the force of dynamical friction is independent of the reference frame.
The torque can then be calculated,
\begin{equation}
    \mathbf{T}_\mathrm{B} = \mathbf{r}_1 \times \mathbf{F}_1 + \mathbf{r}_2 \times \mathbf{F}_2,
\end{equation}
where $\mathbf{F}_i$ is the dynamical friction force with a magnitude given by Eq.~\ref{eq:df-force} and orientated in the opposite direction to $\mathbf{v}_i$. `$\times$' denotes the vector product. The component of $\mathbf{T}_\mathrm{B}$ parallel to the orbital angular momentum is the total torque on the binary orbit,
\begin{equation}
    T_\mathrm{B} = \frac{-2Aa_\mathrm{B}}{c^2} \left(
        \frac{\beta \cos (2\phi) + \cos{\phi}}{f_+ \left[ \mathcal{M}^2 f_+^2 + 4 \right]} +
        \frac{\beta \cos (2\phi) - \cos{\phi}}{f_- \left[ \mathcal{M}^2 f_-^2 + 4 \right]}`
    \right),
\end{equation}
where we have introduced the quantity,
\begin{equation}
    f_\pm = \sqrt{1 + \beta^2 \pm 2\beta \cos \phi}\,.
\end{equation}

The torque on the triple orbit depends on the ratio of the binary separation to its Bondi-Hoyle radius. When the binary separation is shorter than $r_\mathrm{BH}$, the triple torque has a relatively simple form,
\begin{equation}
    T_{\mathrm{T}, a_\mathrm{B} < r_\mathrm{BH}} = \frac{ - 16 A  a_\mathrm{T}}{c^2 \left( \mathcal{M}^2 + 4 \right)}~.
\end{equation}
However, when the binary is wider than its own Bondi-Hoyle radius, its component stars experience dynamical friction individually, and the resulting torque is best expressed as the sum of the contributions from each star,

\begin{equation}
\begin{split}
    &T_1 = \frac{2Aa_\mathrm{B}}{c^2 f_+ \left( \mathcal{M}^2 f_+^2 + 4 \right)}
    \left( \beta \sin^2 \phi + (2\beta^2 + \cos \phi)(1 + \beta\cos\phi) \right) \\
    &T_2 = \frac{2Aa_\mathrm{B}}{c^2 f_- \left( \mathcal{M}^2 f_-^2 + 4 \right)}
    \left( \beta \sin^2 \phi + (2\beta^2 - \cos \phi)(1 - \beta\cos\phi) \right) \\
    &T_3 = \frac{-16 A \beta^2 a_\mathrm{B}}{c^2 \left( \mathcal{M}^2 + 4 \right)}\\
    &T_{\mathrm{T}, a_\mathrm{B} > r_\mathrm{BH}} = T_1 + T_2 + T_3\,,
\end{split}
\end{equation}
where, for consistency, we have expressed $a_\mathrm{T}$ as $2\beta^2 a_\mathrm{B}$.

\section{Orbital evolution under dynamical friction}
\label{app:df-mass-ratio}
Ignoring the phase-dependence of the binary orbit and the geometrical dynamical-friction factor, which later cancels out, and under the assumptions that $v_\mathrm{B} \gg v_{\mathrm{T}}$ and $v_\mathrm{B} \gg c$, the torque on the binary orbit is,
\begin{equation}
    T_\mathrm{B} = - G M_\mathrm{B} \rho_\mathrm{B} a_\mathrm{B}^2\,,
\end{equation}
which can be obtained from Eq.~\ref{eq:binary-torque} in the limit of large $\beta$.
This is an overestimate because it only considers the friction forces at the moment when the binary and triple velocities are parallel.
The corresponding equation for the evolution of the binary orbit is,
\begin{equation}
    \dot{a}_\mathrm{B} = - 8 \rho_\mathrm{B} \sqrt{\frac{G}{M_\mathrm{B}}} a_\mathrm{B}^{5/2} = - \mathcal{B} a_\mathrm{B}^{5/2}\,,
\end{equation}
which defines $\mathcal{B}$. One can derive a similar equation for the triple orbit, with the complication that the two components move through regions with different densities,
\begin{equation}
    \dot{a}_\mathrm{T} = - 2 \rho_\mathrm{3} \sqrt{\frac{G}{M_\mathrm{T}}}
    \left(
    \frac{\rho_\mathrm{B}}{\rho_3} \frac{M_\mathrm{B}}{M_3} \frac{M_\mathrm{T}}{M_3}
    +
    \frac{M_\mathrm{3}}{M_\mathrm{B}} \frac{M_\mathrm{T}}{M_\mathrm{B}}
    \right) a_\mathrm{T}^{5/2} = - \mathcal{T} a_\mathrm{T}^{5/2}\,,
\end{equation}
which defines $\mathcal{T}$.
The corresponding solutions are
\begin{equation}
    a_\mathrm{B}(t) = \left( \tfrac{3}{2} \mathcal{B} t + a_{\mathrm{B}0}^{-3/2} \right)^{-2/3},
    \label{eq:ab-evolution}
\end{equation}
and
\begin{equation}
    a_\mathrm{T}(t) = \left( \tfrac{3}{2} \mathcal{T} t + a_{\mathrm{T}0}^{-3/2} 
    \right)^{-2/3},
    \label{eq:at-evolution}
\end{equation}
where $a_{\mathrm{B}0}$ and $a_{\mathrm{T}0}$ are the initial values of $a_{\mathrm{B}}$ and $a_{\mathrm{T}}$, respectively. Both densities, $\rho_\mathrm{B}$ and $\rho_\mathrm{T}$, are assumed to be constant. This is not physically valid but the assumption does not affect the final result of this section as the varying density amounts to a monotonic remapping of the time coordinate.
Defining $\alpha$ to be the initial ratio of separations, $\alpha = a_{\mathrm{B}0} / a_{\mathrm{T}0}$, one can divide \ref{eq:ab-evolution} by \ref{eq:at-evolution} to determine how this ratio evolves with time:
\begin{equation}
    \frac{a_\mathrm{B}}{a_\mathrm{T}} = \left( \frac{ 3 \mathcal{T}t + 2 a_{\mathrm{T}0}^{-3/2}}  { 3 \mathcal{B} t + 2 \alpha^{-3/2} a_{\mathrm{T}0}^{-3/2} } \right)^{2/3}\,.
\end{equation}

To determine whether the triple reaches dynamical instability, we can set the ratio $a_\mathrm{B}/a_\mathrm{T}$ to an arbitrary value, $x$ (where $\alpha < x < 1$), and solve for $t$,
\begin{equation}
    t = \frac
    { 2 \left( x^{3/2} \alpha^{-3/2} - 1 \right) }
    {3 a_{\mathrm{T}0}^{3/2} \left( \mathcal{T} - \mathcal{B} x^{3/2} \right)}
    \,.
    \label{eq:ratio-crossing-time}
\end{equation}
Owing to the conditions we have imposed on $x$, the numerator of \ref{eq:ratio-crossing-time} is always positive, so the condition for reaching dynamical instability is that the denominator is also positive, i.e.,
\begin{equation}
    x^{3/2} \mathcal{B} < \mathcal{T}\,,
\end{equation}
which is now independent of the initial ratio of separations, $\alpha$.
Upon reversing the substitutions $\mathcal{B}$ and $\mathcal{T}$, this becomes
\begin{equation}
    8 x^{3/2} \sqrt{\frac{G}{M_\mathrm{B}}} \rho_\mathrm{B}
    <
    2 \rho_3 \sqrt{\frac{G}{M_\mathrm{T}}}
    \left(
    \frac{\rho_\mathrm{B}}{\rho_3} \frac{M_\mathrm{B}}{M_3} \frac{M_\mathrm{T}}{M_3}
    +
    \frac{M_\mathrm{3}}{M_\mathrm{B}} \frac{M_\mathrm{T}}{M_\mathrm{B}}
    \right)
\end{equation}
or, upon rearranging,
\begin{equation}
    x^{3/2}
    <
    \frac{1}{4} \sqrt{\frac{M_\mathrm{B}}{M_\mathrm{T}}}
    \left(
    \frac{M_\mathrm{B}}{M_3} \frac{M_\mathrm{T}}{M_3}
    +
    \frac{\rho_3}{\rho_\mathrm{B}} \frac{M_\mathrm{3}}{M_\mathrm{B}} \frac{M_\mathrm{T}}{M_\mathrm{B}}
    \right)\,.
    \label{eq:dyn-inst-condition}
\end{equation}

We must now account for the ratio $\rho_\mathrm{T}/\rho_\mathrm{B}$. If we treat the density of the common envelope as a power law in radius, $\rho \propto r^\gamma$, where $r$ is the distance from the triple's centre of mass, the density ratio can be expressed in terms of the ratio of masses in the system,
\begin{equation}
    \frac{\rho_\mathrm{3}}{\rho_\mathrm{B}} =
    \left( \frac{r_\mathrm{3}}{r_\mathrm{B}}\right)^\gamma =
    \left( \frac{a_\mathrm{T} [M_\mathrm{B} / M_\mathrm{T}]}{a_\mathrm{T} [M_3  / M_\mathrm{T}]} \right)^\gamma = 
    \left(\frac{M_\mathrm{3}}{M_\mathrm{B}}\right)^{-\gamma} = 
    q^{-\gamma}\,,
\end{equation}
where we have introduced the triple mass ratio, $q = M_3/M_\mathrm{B}$. With $M_\mathrm{T}=M_\mathrm{B}+M_{3}$, Eq.~\ref{eq:dyn-inst-condition} becomes,
\begin{equation}
    x^{3/2} < \frac{1}{4} \sqrt{\frac{1}{1+q}}
    \left(
    \frac{1+q}{q^2}
    +
    q^{1-\gamma}[1+q]
    \right)\,,
    \label{eq:q-condition}
\end{equation}
which can be solved numerically for  suitably chosen $x$ and $\gamma$.
For $x \lesssim 0.728$ and $\gamma = -4$, we find that the condition \ref{eq:q-condition} is satisfied for all $q$.
For small $q$, this dependence on $\gamma$ is very weak, meaning that the envelope structure has little influence on whether or not the triple will reach dynamical instability.
When $q$ is large, the envelope structure is more important, and envelopes with steeper density profiles are more likely to produce binary mergers before the three-body instability.


\bsp	
\label{lastpage}
\end{document}